\documentclass[12pt,preprint]{aastex}

\slugcomment{Not to appear in Nonlearned J., 45.}

\shorttitle{Radio Emission from Low-mass SMBH}
\shortauthors{Wang, J et al.}

\begin{document}


\title{Powerful Radio Emission From Low-mass Supermassive Black Holes Favors Disk-like Bulges}


\author{J. Wang\altaffilmark{1,2}, Y. Xu\altaffilmark{1,2}, D. W. Xu\altaffilmark{1,2} and J. Y. Wei\altaffilmark{1,2}}
\email{wj@bao.ac.cn}
\altaffiltext{1}{CAS Key Laboratory of Space Astronomy and Technology, National Astronomical Observatories, Chinese Academy of Sciences}
\altaffiltext{2}{School of Astronomy and Space Science, University of Chinese Academy of Sciences}

%\author{C. D. Biemesderfer\altaffilmark{4,5}}
%\affil{National Optical Astronomy Observatories, Tucson, AZ 85719}
%\email{aastex-help@aas.org}

%\and

%\author{R. J. Hanisch\altaffilmark{5}}
%\affil{Space Telescope Science Institute, Baltimore, MD 21218}


%\altaffiltext{1}{Visiting Astronomer, Cerro Tololo Inter-American Observatory.
%CTIO is operated by AURA, Inc.\ under contract to the National Science
%\altaffiltext{2}{Society of Fellows, Harvard University.}
%\altaffiltext{3}{present address: Center for Astrophysics,
%    60 Garden Street, Cambridge, MA 02138}
%\altaffiltext{4}{Visiting Programmer, Space Telescope Science Institute}
%\altaffiltext{5}{Patron, Alonso's Bar and Grill}


\begin{abstract}

The origin of spin of low-mass \rm supermassive black hole (SMBH) is 
still a puzzle at present.
We here report a study on the host galaxies of a sample of radio-selected nearby ($z<0.05$) Seyfert 2 galaxies
with a BH mass of $10^{6-7} M_\odot$. By modeling the SDSS $r$-band images of these 
galaxies through a 2-dimensional bulge+disk decomposition, we identify a new dependence of
SMBH's radio power on host bulge surface brightness profile, in which
more powerful radio emission comes from a SMBH associated with a more disk-like bulge. 
This result means 
low-mass and high-mass SMBHs are spun up by two entirely different modes that correspond to two different evolutionary paths. 
A low-mass SMBH is spun up by a gas accretion with 
significant disk-like rotational dynamics of the host galaxy in the secular evolution, while a high-mass one by a BH-BH merger in 
the merger evolution.
\end{abstract}



%% tied to the reference list via symbolic KEYs. ../aastex52/The KEY corresponds


\section{INTRODUCTION}

It was for a long time to recognize a remarkable radio-loud/radio-quiet (RL/RQ) dichotomy
for active galactic nuclei (AGNs). RL-AGNs prefer to be associated with high-mass supermassive black holes (SMBHs), i.e., 
$M_{\mathrm{BH}}>10^8 M_{\odot}$ (e.g., Laor 2000).
While, a very wide range of $M_{\mathrm{BH}}$ can be found for RQ-AGNs. 
The RL/RQ dichotomy is, however, challenged by the identification of dozens of RL narrow-line Seyfert 1 galaxies
(e.g., Komossa et al. 2006). The special observational properties of NLS1s 
enable most of authors believe that they are the objects at early evolutionary phase associated 
with low-mass SMBH and very high Eddington ratio (e.g., Mathur 2000; Zhou et al. 2006).

BH's spin is widely believed to play a crucial role in determining the radio emission from SMBHs.
A scenario that an energy extraction from a SMBH spun up by a BH-BH merger 
(e.g., Chiaberge \& Marconi 2011) is favored for high-mass SMBHs. 
This spin-up scenario is, however, almost infeasible for low-mass SMBHs,
because they are believed to largely build from secular evolution driven by internal dynamical processes
in the disk growth rather than from a merge of two BHs (e.g., Kormendy \& Ho 2013).

The origin of angular momentum of these low-mass SMBHs is therefore still a puzzle at present.
In this paper, we attempt to address the puzzle from a different perspective based on the properties of 
the host galaxy that can provide a clue of evolutionary information through stellar dynamics and population.
This aim naturally requires us to focus on radio-selected type II AGNs with small $M_{\mathrm{BH}}$ in stead of 
their type I counterparts, because the obscuration of the central bright nuclei by the dust torus
allows the stellar population and morphology of the host galaxies to be easily measured from ground observations for the type II AGNs . 
In a type I AGN, its host galaxy is typically overwhelmed by the AGN's continuum and broad lines in
optical wavelengths.  
%Possible explanations include: disk broad-line region (BLR) seen almost face on, beaming effect, accretion mode, and 
%BH spin effect (see discussions in Komossa et al. (2006) for a detail).  

%The evidence of beaming effect is identified in some RL-NLS1s,  
%basing upon their both broad band Blazar-like spectral energy distributions 
%(e.g., Zhou et al. 2003, 2005, 2006, 2007; Gallo et al. 2006; Yuan et al. 2008; Abdo et al. 2009); 
%intraday optical variability (Liu et al. 2010, 2016; Maune et al. 2011; Paliya et al. 2013) and high-energy $\gamma$-ray emission with 
%significant variability on a time scale of several days (Abdo et al. 2009a,b,c; Foschini et al. 2010; Eggen et al. 2014;
%Calderone et al. 2011; Paliya et al. 2015). 

%In addition, high-energy $\gamma$-ray emission from jet with 
%significant variability on a time scale of several days 
%is detected in four NLS1s ()
%whose average photon indices are found to be close to that of flat-spectrum radio quasars (Paliya et al. 2015).

%Kormendy \& Kennicutt (2004) suggests that the secular evolution of the disk galaxies results in a pseudobulge with
%a surface-brightness pro?¡§?le close to a classic exponential disk (see also in e.g., Athanassoula 2008; Fisher \& Drory 2011). 
 
The paper is organized as follows. The sample selection and image analysis are described in \S 2 and 3, respectively.
The statistical results along with the implications are shown in \S 4. A $\Lambda$CDM cosmology
with parameters $H\mathrm{_0=70\ km\ s^{-1}\ Mpc^{-1}}$, $\Omega_{\mathrm m}=0.3$, and $\Omega_{\Lambda} = 0.7$ 
is adopted throughout the paper.

\section{SAMPLE SELECTION: SDSS/FIRST NEARBY SEYFERT 2 GALAXIES WITH SMALL BLACK HOLE MASS}

A sample of radio-selected nearby Seyfert 2 galaxies with small $M_{\mathrm{BH}}$ is selected as follows.

\subsection{Seyfert 2 Galaxies with Small $M_{\mathrm{BH}}$}

We start from the value-added SDSS Data Release 7 Max-Planck Institute for 
Astrophysics/Johns Hopkins University (MPA/JHU) catalog (see Heckman \& Kauffmann 2006 for a review).
At the beginning, we require the objects with a redshift smaller than 0.05 to ensure their host galaxies can be 
resolved by the SDSS image observations. Given the line fluxes reported in the catalog, we then
extract ``pure'' Seyfert 2 galaxies from the selected low-z objects
by using a series of demarcation schemes (Kewley et al. 2001, 2006) based on the widely used three 
Baldwin-Phillips-Terlevich diagnostic diagrams (e.g., Veilleux \& Osterbrock 1987).
To further avoid the contamination by low-ionization nuclear emission regions (LINERs), we require the 
[\ion{O}{3}]/[\ion{O}{2}] line ratio\footnote{A correction of local extinction is applied to the observed line fluxes  
by a combination of a Balmer decrement for the standard case B recombination and a
Galactic extinction curve with $R_V=3.1$ throughout the paper.} is larger than 3 according to the 
scheme proposed by Heckman et al. (1981).

Objects with small $M_{\mathrm{BH}}$ are then extracted from the selected ``pure'' Seyfert 2 galaxies
according to the well-documented $M_{\mathrm{BH}}$-$\sigma_\star$ relationship (e.g., Magorrian et al. 1998).
%Gebhardt et al., 2000; Merritt \& Ferrarese 2001; McLure \& Dunlop 2002; Tremaine et al. 2002; Haring \& Rix 2004;
%Ferrarese \& Ford 2005; Aller \& Richstone 2007; Gultekin et al. 2009; Woo et al. 2010; Graham et al. 2011; McConnell \& Ma 2013).  
The values of velocity dispersion $\sigma_\star$ of the star component are taken from the MPA/JHU catalog. 
The objects with measured $\sigma_\star$ between 80 and 120$\mathrm{km\ s^{-1}}$ are retained in our subsequent 
sample selection. This range of $\sigma_\star$ corresponds to a $M_{\mathrm{BH}}$ of $10^{6-7}M_\odot$, based the recent calibration 
of $\log(M_{\mathrm{BH}}/M_\odot)=(8.32\pm0.05)+(5.64\pm0.32)\log(\sigma_\star/200\ \mathrm{km\ s^{-1}})$ presented in 
McConnell \& Ma (2013). The calibration is valid for a sample of $M_{\mathrm{BH}}$ of $10^{6-10}M_\odot$.
%although the $M_{\mathrm{BH}}$-$\sigma_\star$ correlation is argued against by Kormendy et al. (2011) for pseudobulges associated 
%with a small BH (see also in Kormendy \& Ho 2013). 
The lower limit of $\sigma_\star=80\mathrm{km\ s^{-1}}$  is used in our sample selection because 
of the instrumental spectral resolution of SDSS of $\sim70\mathrm{km\ s^{-1}}$.

\subsection{Cross-match with First Survey}

The selected ``pure'' Seyfert 2 galaxies with small $M_{\mathrm{BH}}$ are subsequently
cross-matched with the FIRST survey catalog (Becker et al. 2003). The cross-match follows the methods described in 
Richards et al (2002), in which a matching radius of 2\arcsec\ is adopted for compact radio sources, and a radius of 
$6\arcmin$ for possible extended sources\footnote{At $z=0.01$, the radius of $6\arcmin$ corresponds to a physical size of $\sim70 \mathrm{kpc}$ that 
is close to the size ($\sim100\ \mathrm{kpc}$) of typical jets and lobs. 
In fact, all the objects listed in our final sample have redshifts larger than 0.01 (see Table 1).}. 
%At redshift of 0.01\footnote{Only a few of objects in the input sub-sample virtually 
%have redshifts below this value.}, the radius of $6\arcmin$ corresponds to a physical size of $\sim70 \mathrm{kpc}$ that 
%is close to the size ($\sim100\ \mathrm{kpc}$) of typical jets and lobs.  
With the FIRST limiting flux density (5$\sigma$) of 1mJy, our cross-match finally 
returns 54 radio-selected Seyfert 2 galaxies with small $M_{\mathrm{BH}}$.

The radio power at 1.4 GHz (rest frame) of each of the 54 selected objects is
calculated from the observed integrated flux density $f_\nu$ through the formula
$P_{\mathrm{1.4GHz}}=4\pi d_L^2f_\nu(1+z)^{-1-\alpha}$, where $d_L$ is the
luminosity distance, $z$ the redshift, and $\alpha=-0.8$ (e.g., Ker et al. 2012) the
spectral slope defined as $f_\nu\propto\nu^{\alpha}$. The derived $P_{\mathrm{1.4GHz}}$ has a 
range from $10^{21}$ to $10^{23} \mathrm{W\ Hz^{-1}}$.    
We estimate the radio loudness of each object by using the 
[\ion{O}{3}] luminosity $L_{\mathrm{[OIII]}}$ as a proxy of AGN's bolometric luminosity $L_{\mathrm{bol}}$.   
The combination of the widely used bolometric corrections of $L_{\mathrm{bol}}\approx3500L_{\mathrm{[OIII]}}$ 
and $L_{\mathrm{bol}}=9\lambda L_\lambda(5100\AA)$
(Heckman \& Best 2014; Kaspi et al. 2000) leads to an estimation of radio loudness $R'$
\footnote{
%We here calculate the radio loudness by using the radio flux at 1.4GHz instead of at 5GHz that is adopted 
%by Kellermann et al. (1989). Our definition allows the calculated radio loudness avoids the uncertainty caused 
%by the large dispersion of the spectral slope $\alpha$ (e.g., Ker et al. 2012). 
Assuming an universal spectral slope of $\alpha=-0.8$ yields a transformation of 
$R'_{\mathrm{1.4GHz}}=2.77R_{\mathrm{5GHz}}$. The tiny difference between the monochromatic optical luminosities at 4400\AA\ and 5100\AA\ is ignored in
our study.}
\begin{equation}
  \log R'=\log\bigg(\frac{P_{\mathrm{1.4GHz}}/\mathrm{W\ Hz^{-1}}}{L_{\mathrm{[OIII]}}/\mathrm{erg\ s^{-1}}}\bigg)+19.18
\end{equation}
It must be stressed that the possible systematics of the used bolometric corrections mean that the 
derived $R'$ is only meaningful for a comparison study.

\subsection{Sample Selection on AGN's Activity}

The aim of this study is to explore the effect of host galaxy on the origin of radio emission in AGNs,
which requires a sample selected on nuclear accretion property.   
Figure 1 shows an anti-correlation between the calculated $L_{\mathrm{[OIII]}}$ and $R'$ for all the selected 
54 objects. The anti-correlation motivates us to exclude the objects at the 
right-bottom (left-top) corner for the current study, because 
the most large (small) $R'$ of these objects are simply caused by their extremely low (high) accretion activities rather than 
powerful (weak) radio emission. With these considerations, we finally focus on the objects located within a bin of
$\log L_{\mathrm{[OIII]}}=40.6-41.2$. The bin size is chosen by a compromise between 
the scatter of the $L_{\mathrm{[OIII]}}$ versus $R'$ correlation and the size of our finally used sample.

\vspace{0.2cm}

In summary, we finally selected 31 radio-selected nearby ($z<0.05$) Seyfert 2 galaxies with 
small $M_{\mathrm{BH}}$ of $10^{6-7}M_\odot$, after removing the duplications. The properties of these objects are 
listed in Table 1, except for SDSS\,J160151.51+024809.9 (see Section 3 for the details). 
%Columns (1) and (2) list the identification of each object and the corresponding
%redshift given by the SDSS pipelines (Glazebrook
%et al. 1998; Bromley et al. 1998). The derived [\ion{O}{3}] line luminosity and radio power at 1.4GHz are 
%tabulated in Column (3) and (4), respectively. The column (5) presents the calculated radio loundness according to 
%Eq. (1). The correspond errors are obtained by a proper error propagation.
%The $M_{\mathrm{BH}}$ estimated from stellar velocity dispersion by the calibration given in McConnell \& Ma (2013) is
%listed in Column (6). 
%Column (7) lists the 4000\AA\ break index $D_n(4000)$ (Bruzual 1983; Balogh et al. 1999) 
%defined as $D_n(4000)=\int_{4000}^{4100} f_\lambda d\lambda/\int_{3850}^{3950} f_\lambda d\lambda$, 
%taken from the MPA/JHU catalog. The index
%is popularly used as an excellent mean age indicator of the stellar
%population of the bulge of a galaxy until a few Gyr after the onset
%of a star formation activity (e.g., Bruzual \& Charlot 2003; Heckman \& Kauffmann 2006; Coelho et al. 2007).

\section{DATA REDUCTIONS: TWO-DIMENSIONAL BULGE/DISK DECOMPOSITION}

At first, the SEXTRACTOR package (Bertin \& Arnouts 1996) is 
performed for each of the SDSS-corrected $r$-band frames to determine the detection threshold, and to define the 
area where the galaxy signal is above the determined threshold. 
A threshold of 1.5 times of the background noise is adopted in our data reduction. 
A stamp image centered on the object is therefore 
produced by the SEXTRACTOR package.
With the stamp images, we perform a 2-dimensional bulge+disk decomposition    
by the publicly available GIM2D package\footnote{The home page of the GIM2D package can be found at 
http://astrowww.phys.uvic.ca/~simard/GIM2D/.} (Simard et al. 2002) that is demonstrated to be valid for 
different galaxy samples (e.g., de Jong et al. 2004), 
except for SDSS\,J160151.51+024809.9. The image shows that
the host of the object is a heavily obscured edge-on disk galaxy. The surface
brightness profile used in our decomposition is described by a combination of an exponential radial profile for the disk component
and a Sersic profile with an index of $n_\mathrm B$ for the bulge component.  
A simple point spread function (PSF) with a Gaussian profile is adopted in the convolution of our modeling to account for the seeing effect.   
The resulted reduced $\chi^2$ is very close to unit for all the 30 host galaxies.
 
Table 1 tabulates some modeled parameters of the 30 host galaxies. The parameters 
are B/T ratio, the ratio of the disk scale length over the bulge effective radius $h_d/r_e$ and the Sersic index
$n_\mathrm B$. The reported uncertainties given by the GIM2D 
package at a confidence level of 99\% are based on the topology of the parameter space being built up in the fitting. 

\section{RESULTS AND IMPLICATIONS}

%The aim of this paper is to explore the origin of the intensive radio emission from less massive SMBH, such as the cases of the RL-NLS1s,
%by using a sample of radio-selected nearby Seyfert 2 galaxies associated a SMBH within the mass range of $10^{6-7}M_\odot$. 
Two objects (SDSS\,J154559.09+270629.5 and SDSS\,J162622.65+210542.8)  
are excluded from subsequent statistical analysis because 
their modeled bulge effective radius $r_e$ are extremely smaller than the size of the corresponding PSF. 
%We clarify that the exclusion has negligible influence on our final statistical results.     

\subsection{Statistical Results}

The modeled $n_\mathrm{B}$ is plotted against the calculated $R'$, 
$L_{\mathrm{[OIII]}}$, and $P_{\mathrm{1.4GHz}}$ in the left, middle and right panels in Figure 5, respectively.
The left panel shows a newly identified anti-correlation between $R'$ and $n_\mathrm{B}$.
%One can see from the plot that 
%stronger the radio emission with respect to the AGN's accretion emission, smaller the Sersic index (i.e., more likely a exponential bulge)
%will be. 
In fact, the galaxies with $R'<1$ generally tend to have a classical bulge with $n_\mathrm B>2.0$, while the ones 
with $R'>1$ a pesudo-bulge with $n_\mathrm B<2.0$ (e.g., Kormendy \& Kennicutt 2004; Fisher \& Drory 2008). 
A Spearman rank-order statistical test yields a correlation coefficient of $r_s=-0.54$. The corresponding probability of
null correlation from two-tailed is calculated to be $p_s=0.0029$, which corresponds to a significant level at 2.81$\sigma$.
As an additional test, a generalized  Kendall's  $\tau$ correlation coefficient is calculated to be 
$\tau=-0.78$ at a significance level of $2.91\sigma$. The corresponding probability of null correlation is inferred to be 
$p_s=0.0037$. A comparison of the plots in the middle and right panels clearly shows that it is the radio power, 
rather than the AGN's accretion activity, depending on the modeled $n_{\mathrm B}$.

Figure 6 plots $R'$ against various properties of host galaxy.
At first, one can see from the plot that there is an independence of $R'$ on the stellar 
population age as assessed by the parameter $D_n(4000)$, although almost all the host galaxies are 
associated with relatively young stellar populations , i.e., $D_n(4000)<1.6$ (e.g., Heckman \& Best 2014). 
%There are only two 
%objects with very low $R'$ have passively evolved host galaxies with $D_n(4000)>1.7$ .  
Another independence can be identified for $R'$ on the measured bulge fraction B/T, which implies that the radio emission from the
low-mass SMBHs is not related with the morphology type of the host galaxies. 
%$R'$ is plotted against the ratio of the disk scale length 
%over the bulge effective radius $h_d/r_e$ in the bottom-right panel, which enables us to 
We identify a moderate relation between $R'$ and $h_d/r_e$. Objects with stronger radio emission tends to have 
larger bulge size. Our statistical analysis returns a correlation coefficient of $r_s=-0.48$ at a significance level 
of $2.47\sigma$ by adopting the Spearman rank-order statistics. The corresponding probability of
null correlation is determined to be $p_s=0.0106$. In addition, a correlation coefficient based on the 
Kendall's $\tau$ method is obtained to be $\tau=-0.65$ at a significance level of $2.41\sigma$, which 
corresponds to a probability of null correlation of $p_s=0.0159$.

\subsection{Implications}

One would argue that the powerful radio emission in the sample is emitted from starforming regions in the host galaxies rather than 
from the AGNs (e.g., Kennicutt 1992).
We estimate an upper limit of star formation rate (SFR) for each object
through the calibration of $\mathrm{SFR}=7.9\times\frac{L_{\mathrm{[OII]}}/10^{42} \mathrm{erg\ s^{-1}}}{16.73-1.75[\log(\mathrm{O/H})+12]} M_\odot\ \mathrm{yr^{-1}}$
in Kewley et al. (2004), by assuming the [OII]$\lambda3727$ line emission is fully contributed by the ongoing 
star formation. $\log(\mathrm{O/H})+12=9.2$ is the metallicity 
twice of the solar value that is usually used in AGNs (e.g., Ho 2005).
With the estimated SFR, an upper limit of the radio power 
contributed by an underlying star formation $P_{\mathrm{exp}}$ is then
inferred for each object by a combination of the relationship
of $\mathrm{SFR}(M\geq5M_\odot)=\frac{P_{\mathrm{exp,1.4GHz}}}{4.0\times10^{21}\mathrm{W\ Hz^{-1}}} M_\odot\ \mathrm{yr^{-1}}$ 
in Condon (1992)
%\begin{equation}
%$\mathrm{SFR}(M\geq5M_\odot)=\frac{P_{\mathrm{exp,1.4GHz}}}{4.0\times10^{21}\mathrm{W\ Hz^{-1}}} M_\odot\ \mathrm{yr^{-1}}$ 
%\end{equation}
and a Salpeter initial mass function.
The bottom-left panel 
in Figure 6 shows the distribution of $P_{\mathrm{obs}}/P_{\mathrm{exp}}$, where $P_{\mathrm{obs}}$ is 
the observed radio power. 
The distribution indicates that the values of $P_{\mathrm{exp}}$ are typically
lower than the observed ones by 1-2 orders of magnitudes, with the worst case of $P_{\mathrm{exp}}=0.25P_{\mathrm{obs}}$.    

Previous studies have reported two possible problems with the $M_{\mathrm{BH}}$ estimated from the $M_{\mathrm{BH}}-\sigma_\star$ correlation 
for pesudo-bulges. The relationship of pesudo-bulge is either very weak (e.g., Kormendy et al. 2011) or 
shifted from that of classical bulge in both slope and intercept (e.g., 
Hu 2008; McConnell \& Ma 2013; Ho \& Kim 2014). These two problems have no essential effect 
on our results. At first, the values of $M_{\mathrm{BH}}$ listed in Table 1 are only representative in sample description and not directly involved
in the revealed $R'-n_{\mathrm{B}}$ anti-correlation. Secondly,
compared to the relationship of pesudo-bulges,
the $M_{\mathrm{BH}}$ estimated from the global relationship tends to be systematically over-estimated, rather than under-estimated,
which in fact reinforces our sample selection of objects with small $M_{\mathrm{BH}}$.      

%compared to the relationship of late-type galaxies given in McConnell \& Ma (2013), 
%the $M_{\mathrm{BH}}$ estimated from the global relationship is systematically over-estimated, rather than under-estimated, by a factor 
%of 1.3 for a $\sigma_\star=120\ \mathrm{km\ s^{-1}}$, which in fact reinforces our sample selection of objects with small $M_{\mathrm{BH}}$
%(a similar conclusion can be obtained from the results in Ho \& Kim 2014).    

\rm

Both orientation and intrinsic mechanisms have been proposed to explain the powerful radio emission 
in AGNs with small $M_{\mathrm{BH}}$. The orientation mechanism attributes the observed 
powerful radio emission to either a significant underestimation of $M_{\mathrm{BH}}$ due to a disk broad-line region (BLR) almost face-on
(e.g., Baldi et al. 2016) or 
a boost in radio flux due to the beaming effect of a relativistic jet (e.g., Yuan et al. 2008; Liu et al. 2016 and references therein). 
In the intrinsic mechanism,
the powerful radio emission can be ascribed to an energy extraction from either 
a disk wind (e.g., Blandford \& Payne 1982; Wang et al. 2003; Cao 2016) or BH spin (e.g., Blandford \& Znajek 1977). 

The identified correlation between radio power and host bulge profile at first allows us 
to exclude the orientation scenarios for the current sample, simply because the 
bulge profile is not related with the orientation of either a disk BLR or a relativistic jet. 
In addition,
our sample is selected on $M_{\mathrm{BH}}$ estimated from the $M_{\mathrm{BH}}-\sigma_\star$ relation,
which allows us to completely avoid the underestimation issue mentioned above.

We subsequently argue that the disk wind scenario is not applicable in the current sample, because
our sample is selected on nuclear properties (i.e., accretion rate and $M_{\mathrm{BH}}$).
%, which means an independence of the 
%radio emission on the accretion property. 
In fact, a direct statistical analysis does not reveal any 
relation between $R'$ and Eddington ratio $\lambda_{\mathrm{Edd}}$ assessed by $\lambda_{\mathrm{Edd}}\sim L_{\mathrm{[OIII]}}/\sigma_{\star}^4$
that is usually used in type II AGNs (e.g., Heckman \& Best 2014).

With the exclusion of the three possible scenarios, we finally argue that the $R'-n_{\mathrm B}$ anti-correlation favors the scenario in which a  
low-mass SMBH is spun up by the gas accreted with significant disk-like rotational dynamics.
An extraction of rotational energy of a BH plays an important role in launching powerful jets. 
%BH spin plays an important role in launching powerful jets from a SMBH. 
%Even though the physical details is far from being understood, 
%The emergence of the powerful jets 
%is believed to be relate to the extraction of rotational energy of the BH. 
The dependence of jet power on BH spin has been frequently suggested by various models (e.g., Ghosh \& Abramowicz 1997; Wang et al. 2003), 
although the details depends on the adopted accretion disk model 
and the parametrization of the poloidal magnetic filed. 
%(since the jet power is expected to be\cite{mei01} 
%$L_{\mathrm{jet}}\propto B_{\mathrm p}^2$).
Both BH-BH merger and disk accretion can efficiently shape the final BH spin (e.g., Hughes \& Blandford 2003; Volonteri et al. 2005).
%(egHughes & Blandford 2003; Volonteri et al2005)
The revealed independence of $R'$ on B/T ratio at first allows us to exclude the 
merger scenario for spinning-up the low-mass SMBHs. 

A host rotational dynamics-related BH spinning-up is 
then suggested by the revealed $R'-n_{\mathrm{B}}$ correlation.
A clear bulgeless disk with $n_{\mathrm{B}}\sim1$ is recently discovered in 2MASX\,J23453268-0449256, an unique RL massive spiral galaxy with 
a powerful jet of a scale of $\sim$1.6 Mpc (Bagchi et al. 2014). 
In fact, BH spin is suggested to relate with host dynamics by a recent theoretical study, in which a BH fueling flow
attached to the dynamics of host galaxy at large scale is required to match the observed
BH spin distribution well (Sesana et al. 2014).
Observations indicate that the dynamics in pesudo-bulges is more dominated by rotation
than in classical bulges (Kormendy \& Kennicutt 2004).
%between the bulge surface-brightness profile and radio power. 
A  low-mass  SMBH can be efficiently spun up by the Bardeen-Peterson effect (Bardeen \& Peterson 1975)  
that realign the BH-disk system through the interaction between the frame-dragging effect of the Lense-Thirring torque and the 
strong disk viscous strees (e.g., King et al 2005; Perego et al 2009; Li et al 2015),
if the gas accreted onto the SMBH has not only significant disk-like rotational dynamics,
but also a mass exceeding the alignment mass limit of accretion event.
The alignment mass limit is roughly 
$m_{\mathrm{align}}\sim a M_{\mathrm{BH}}\sqrt{R_\mathrm s/R_\mathrm w}$ (Sikora et al. 2007),
where $a$ is the dimensionless angular momentum 
%$a=J/J_{\mathrm{max}}=cJ/GM^2_{\mathrm{BH}}$
$a=cJ/GM^2_{\mathrm{BH}}$, and 
$R_\mathrm s=2GM_{\mathrm{BH}}/c^2$ and $R_\mathrm w$ are the Schwarzschild radius and distance of wrap of the accretion disk, respectively.  
With the Eq. (22) in King et al. (2005) for $R_\mathrm s/R_\mathrm w$, the limit mass is reduced to 
$m_{\mathrm{align}}\propto a^{11/16}(L/L_{\mathrm{Edd}})^{1/8}M^{15/16}_{\mathrm{BH}}$, which implies that, with 
respect to a high-mass one, a low-mass SMBH can be more readily spun up by a smaller mass increment.

Our results show that low-mass and high-mass SMBHs are spun up by two entirely different
mechanisms that are believed to relate with two evolutionary paths.
A pesudo-bulge that is favored by a spinning-up of low-mass SMBHs 
%with a surface-brightness profile close to a classic exponential disk
can be resulted from the secular evolution of disk galaxies (e.g, Silverman et al. 2011; 
Kormendy \& Ho 2013; Kormendy \& Kennicutt 2004; Fisher \& Drory 2011), in which the 
pesudo-bulge might be related with either a second hump instability or a vertical dynamical resonance (e.g., Sellwood 2014).
Meanwhile, a classical bulge favored for high-mass SMBHs is widely believed to be created through a ``dry'' merger of two
galaxies (Toomre 1977),although this scenario is challenged by a recent identification of a RL and bulgeless spiral galaxy
with a $M_{\mathrm{BH}}\sim10^8M_\odot$ (Bagchi et al. 2014).

\section{CONCLUSION}

We study the origin of spin of small SMBH on   
a sample of radio-selected nearby ($z<0.05$) Seyfert 2 galaxies
with a $M_{\mathrm{BH}}$ of $10^{6-7} M_\odot$, which 
allows us to identify a new dependence of radio power on host bulge surface brightness.
The dependence favors a scenario that a low-mass SMBH is spun up by the gas accreted with 
significant disk-like rotational dynamics.

\acknowledgments

The author would like to thank the anonymous referee for his/her very useful
comments and suggestions for improving the manuscript. 
%The authors thank the anonymous referee for his/her careful review and helpful suggestions
%for improving the manuscript. We thanks Dr. X. L. Zhou for the help in X-ray spectral analysis.
This study uses the
SDSS archive data that was created and distributed by the Alfred P. Sloan Foundation.
%This work is based on observations obtained with XMM-Newton, an ESA science mission with
%instruments and contributions directly funded by ESA Member
%States and the USA (NASA).
The study is supported by the National Basic Research Program of China (grant
2009CB824800) and by National Natural Science
Foundation of China under grants 11473036 and 11273027.

\clearpage

\begin{table}
\begin{center}
\caption{Properties of the 30 radio-selected nearby ($z<0.05$) Seyfert 2 galaxies with small $M_{\mathrm{BH}}$.\label{tbl-2}}
\tiny
\begin{tabular}{cccccccccc}
\tableline\tableline
SDSS  & z & $L_{\mathrm{[OIII]}}$  & $P_{\mathrm{1.4GHz}}$ & $\log R'$ & $\log(M_{\mathrm{BH}}/M_\odot)$ & $D_n(4000)$ & B/T & $h_d/r_e$ & $n_{\mathrm{B}}$ \\
      &   &   $10^{40}\ \mathrm{erg\ s^{-1}}$ & $10^{21}\ \mathrm{W\ Hz^{-1}}$ &  &  &  &  &  \\
(1) & (2) & (3) & (4) & (5) & (6) & (7) & (8) & (9) & (10)\\ 
\tableline
J005329.92-084604.0   &  0.0190 & $20.37\pm0.15$  &     $5.68\pm0.12$   & $-0.37\pm0.02$ &   6.84     &   1.37      & $0.58\pm0.01$ & $1.30\pm0.08$ & $1.40\pm0.02$\\
J024703.64-003533.2   &  0.0423 &  $7.30\pm0.12$  &    $10.96\pm0.57$   & $0.36\pm0.05$ &   6.72     &   1.47       & $0.70\pm0.02$ & $0.35\pm0.15$ & $1.88\pm0.07$\\
J025329.59-001405.5   &  0.0288 & $20.78\pm0.12$  &    $13.00\pm0.20$   & $-0.02\pm0.02$ &   6.52     &   1.43      & $0.54\pm0.03$ & $2.24\pm0.44$ & $1.62\pm0.05$\\
J073715.73+313110.9   &  0.0269 & $13.44\pm0.05$  &     $3.30\pm0.22$   & $-0.43\pm0.07$ &   7.00     &   1.74      & $0.53\pm0.02$ & $2.09\pm0.47$ & $2.95\pm0.12$\\
J080547.34+225434.8   &  0.0304 & $25.48\pm0.07$  &    $11.29\pm0.27$   & $-0.17\pm0.02$ &   6.96     &   1.36      & $0.26\pm0.01$ & $3.30\pm0.34$ & $1.04^{+0.06}_{-0.04}$\\
J082443.28+295923.5   &  0.0254 & $28.77\pm0.18$  &     $2.56\pm0.21$   & $-0.87\pm0.08$ &   6.78     &   1.30      & $0.37\pm0.01$ & $10.41\pm1.11$ & $2.48\pm0.03$\\
J090613.76+561015.2   &  0.0466 &  $7.39\pm0.07$  &    $23.36\pm0.79$   & $0.68\pm0.04$ &   6.90     &   1.26       & $0.48\pm0.07$ & $4.50\pm1.90$ & $2.24\pm0.08$\\
J093236.58+095025.9   &  0.0489 & $18.36\pm0.11$  &    $14.55\pm0.83$   & $0.08\pm0.06$ &   6.83     &   1.39       & $0.06\pm0.01$ & $6.30\pm2.08$ & $2.42\pm0.03$\\
J094044.50+211403.3   &  0.0244 & $15.52\pm0.10$  &     $1.50\pm0.18$   & $-0.84\pm0.12$ &   6.46     &   1.50      & $0.24\pm0.01$ & $2.02\pm0.23$ & $3.06\pm0.08$\\
J095742.84+403315.8\tablenotemark{a}   &  0.0453 &  $7.26\pm0.08$  &    $10.58\pm0.65$   & $0.34\pm0.06$ &   6.75     &   1.53      & $0.35\pm0.02$ & $1.87\pm0.32$ & $1.00^{+0.01}_{-0.00}$\\
J112008.68+341845.8   &  0.0367 &  $6.34\pm0.08$  &     $2.60\pm0.41$   & $-0.21\pm0.16$ &   6.66     &   1.53      & $0.39\pm0.03$ & $0.38\pm0.21$ & $1.84\pm0.07$\\
J112135.17+042647.2   &  0.0470 &  $4.46\pm0.09$  &     $5.10\pm0.76$   & $0.24\pm0.15$ &   6.63     &   1.49       & $0.43\pm0.05$ & $0.66\pm0.34$ & $2.24\pm0.09$\\
J113630.49+265138.8\tablenotemark{b}   &  0.0333 & $19.42\pm0.10$  &    $28.95\pm0.36$   & $0.35\pm0.01$ &   6.42     &   1.37      & $0.20\pm0.00$ & $2.48\pm0.55$ & $4.00\pm0.00$\\
J113808.01+111146.9\tablenotemark{a}   &  0.0357 &  $8.01\pm0.08$  &    $25.02\pm0.42$   & $0.67\pm0.02$ &   6.87     &   1.48       & $0.00\pm0.00$ & \dotfill & $1.00\pm0.00$\\
J114216.87+140359.7   &  0.0207 & $20.16\pm0.07$  &     $1.76\pm0.13$   & $-0.88\pm0.07$ &   6.29     &   1.51      & $0.21\pm0.01$ & $8.69\pm0.41$ & $3.31\pm0.06$\\
J115429.40+425848.6   &  0.0235 &  $7.06\pm0.05$  &     $1.39\pm0.17$   & $-0.53\pm0.13$ &   6.20     &   1.53      & $0.26\pm0.01$ & $5.19\pm0.55$ & $2.59\pm0.16$\\
J122119.70+544923.2   &  0.0375 &  $9.59\pm0.08$  &     $9.72\pm0.46$   & $0.19\pm0.05$ &   6.58     &   1.34       & $0.80\pm0.04$ & $2.54\pm0.57$ & $3.48\pm0.19$\\
J122438.68+013243.0   &  0.0256 &  $6.54\pm0.06$  &     $1.81\pm0.23$   & $-0.38\pm0.13$ &   6.84     &   1.43      & $0.21\pm0.01$ & $4.68\pm0.29$ & $3.88\pm0.12$\\
J124054.96+080323.2\tablenotemark{a}   &  0.0478 & $23.39\pm0.15$  &     $9.78\pm0.72$   & $-0.20\pm0.07$ &   6.55     &   1.41      & $0.28\pm0.06$ & $0.61\pm0.20$ & $1.00\pm0.00$\\
J140040.56-015518.2\tablenotemark{b}   &  0.0250 & $25.54\pm0.12$  &     $3.78\pm0.21$   & $-0.65\pm0.06$ &   6.69     &   1.19      & $0.20\pm0.01$ & $1.45\pm0.29$ & $4.00^{+0.00}_{-0.01}$\\
J140804.00+071939.5   &  0.0238 &  $4.78\pm0.05$  &     $3.43\pm0.19$   & $0.04\pm0.06$ &   6.13     &   1.26       & $0.43\pm0.03$ & $6.44\pm1.53$ & $2.22\pm0.07$\\
J141041.34+133328.7   &  0.0162 & $11.04\pm0.09$  &     $5.56\pm0.09$   & $-0.12\pm0.02$ &   6.31     &   1.36      & $0.37\pm0.01$ & $4.96\pm0.21$ & $3.10\pm0.10$\\
J151621.58+080604.7\tablenotemark{a}   &  0.0309 & $14.66\pm0.14$  &    $31.12\pm0.29$   & $0.51\pm0.01$ &   6.83     &   1.46       & $0.28\pm0.01$ & $0.40\pm0.18$ & $1.00^{+0.03}_{-0.00}$\\
J153926.06+245636.8\tablenotemark{a}   &  0.0228 & $11.96\pm0.06$  &    $10.22\pm0.17$   & $0.11\pm0.02$ &   6.54     &   1.18       & $0.39\pm0.01$ & $1.58\pm0.06$ & $1.00^{+0.02}_{-0.00}$\\
J154304.09+511248.9   &  0.0362 & $14.68\pm0.10$  &     $7.77\pm0.41$   & $-0.10\pm0.05$ &   6.16     &   1.30      & $0.44\pm0.02$ & $0.56\pm0.21$ & $2.77\pm0.19$\\
J154559.09+270629.5   &  0.0314 &  $5.34\pm0.08$  &     $3.97\pm0.32$   & $0.05\pm0.08$ &   6.47     &   1.49       & $0.03\pm0.01$ & $67.93\pm7.21$ & $2.45\pm0.01$\\
J162451.25+192535.7   &  0.0359 & $15.70\pm0.10$  &     $7.96\pm0.42$   & $-0.16\pm0.05$ &   6.96     &   1.41      & $0.49\pm0.02$ & $2.73\pm0.62$ & $1.99\pm0.06$\\
J163040.91+302919.4   &  0.0368 & $16.83\pm0.12$  &     $6.74\pm0.44$   & $-0.22\pm0.07$ &   7.02     &   1.56      & $0.28\pm0.01$ & $52.90\pm5.94$ & $1.79\pm0.04$\\
J162622.65+210542.8   &  0.0320 &  $5.15\pm0.08$  &     $2.83\pm0.48$   & $-0.08\pm0.17$ &   6.61     &   1.49      & $0.21\pm0.01$ & $3.09\pm0.66$ & $2.65\pm0.04$\\
J215259.07-000903.4   &  0.0277 & $30.59\pm0.05$  &     $1.74\pm0.17$   & $-1.06\pm0.10$ &   7.01     &   1.71      & $0.21\pm0.01$ & $7.02\pm0.70$ & $3.98\pm0.02$\\
                                                      
\tableline
\tablenotetext{a}{The Sersic index is fixed to be the minimum value of $n_{\mathrm B}=1$ for the best-fit model.}
\tablenotetext{b}{The Sersic index is fixed to be the maximum value of $n_{\mathrm B}=4$ for the best-fit model.}

\end{tabular}
\end{center}
\tablecomments{Column (1): SDSS identification. Column (2): Redshift given by the SDSS spectroscopy pipelines. 
Column (3): [OIII]$\lambda5007$ line luminosity in unit of $10^{40}\ \mathrm{erg\ s^{-1}}$. 
Column (4): Radio power at 1.4GHz in unit of $10^{21}\ \mathrm{W\ Hz^{-1}}$. 
Column (5): Radio loudness $R'$ calculated from Eq (1). Column (6): The BH mass estimated from the star light
velocity dispersion $\sigma_\star$ through the well-established $M_{\mathrm{BH}}-\sigma_\star$ relation. 
Column (7): Parameter of 4000\AA\ break index 
defined as $D_n(4000)=\int_{4000}^{4100} f_\lambda d\lambda/\int_{3850}^{3950} f_\lambda d\lambda$ (e.g., Coelho et al. 2007). 
%The index is popularly used as an excellent mean age indicator of the stellar
%population of the bulge of a galaxy until a few Gyr after the onset
%of a star formation activity (e.g., Bruzual \& Charlot 2003; Heckman \& Kauffmann 2006; Coelho et al. 2007).. 
Column (8): Bulge fraction obtained by 2-dimensional bulge+disk 
decompositions. Column (9): Ratio of the disk scale length over the bulge effective radius. Column (10): 
Modeled Sersic index. All the errors in Columns (7)-(9) are taken from the results reported by the 
GIM2D package that provides upper
and lower limits of each free parameter at a confidence level of 99\% based on the topology of the
parameter space being built up in the fitting.   
}
\end{table}

%% the file, uncomment the \plotone, \plottwo, and
%% If you need a layout that cannot be achieved with \plotone or
%% \plottwo, you can invoke the graphicx package directly with the
%% \includegraphics command or use \plotfiddle. For more information,
%% please see the tutorial on "Using Electronic Art with AASTeX" in the
%% documentation section at the AASTeX Web site,
%% http://www.journals.uchicago.edu/AAS/AASTeX.
%%
%% The examples below also include sample markup for submission of
%% supplemental electronic materials. As always, be sure to check
%% the instructions to authors for the journal you are submitting to
%% for specific submissions guidelines as they vary from

%% This example uses \plotone to include an EPS file scaled to
%% 80% of its natural size with \epsscale. Its caption
%% has been written to indicate that additional figure parts will be
%% available in the electronic journal.

\begin{figure}
\epsscale{.80}
\plotone{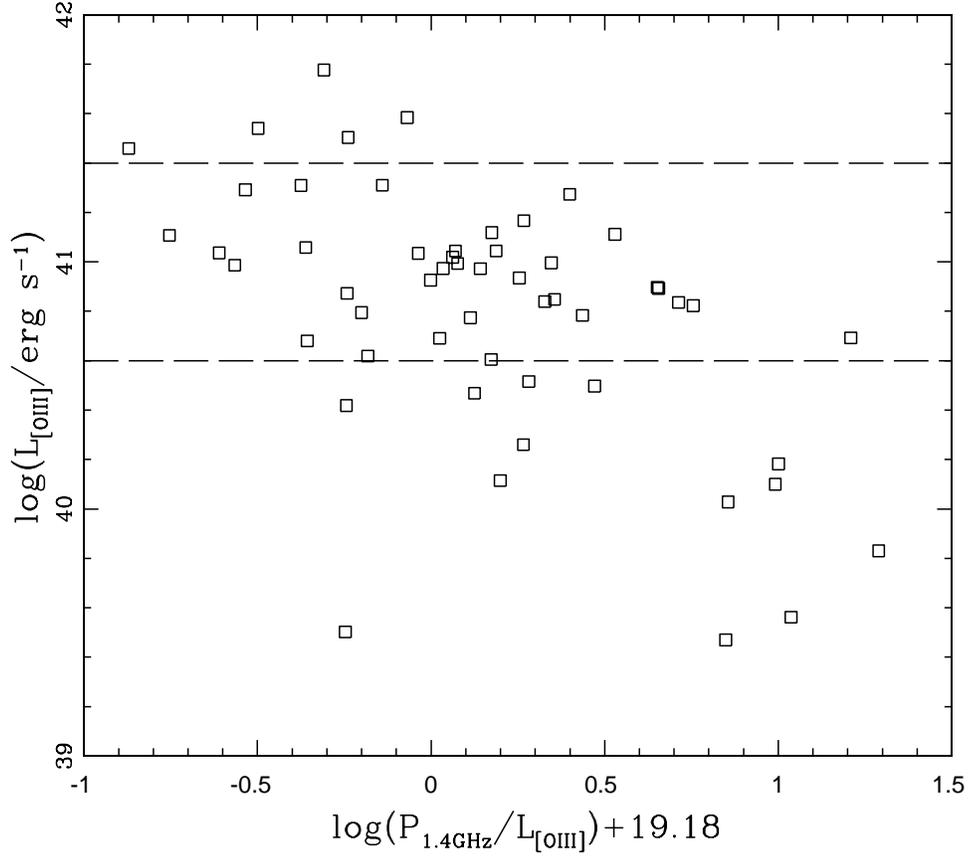}
\caption{[\ion{O}{3}] line luminosity plotted against the estimated radio loudness (see Eq. 1) for the 54 radio-selected nearby Seyfert 2 galaxies 
with small BH mass.  
In order to select a sample on AGN's accretion property, only the 31 objects with $L_{\mathrm{[OIII]}}$ within the 
range (i.e., $\log L_{\mathrm{[OIII]}}=40.6-41.2$) marked by the two dashed lines are considered in our 2-dimensional bulge+disk decompositions.   
}
\end{figure}

\begin{figure}
\epsscale{.80}
\plotone{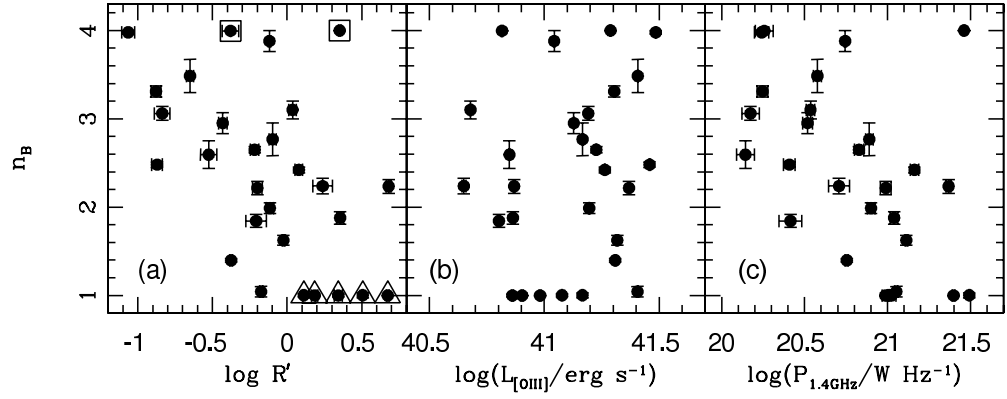}
\caption{\it Left panel: \rm An anti-correlation between the modeled Sersic index $n_\mathrm{B}$ and radio loudness $R'$ estimated from Eq. (1). The objects
with fixed value of $n_\mathrm{B}$ are marked by triangles for $n_\mathrm{B}=1$ and by squares for $n_\mathrm{B}=4$. 
\it Middle panel: \rm $R'$ plotted against [\ion{O}{3}] line luminosity. \it Right panel: \rm The same as the middle one but for 
radio power at 1.4GHz.
}
\end{figure}

\begin{figure}
\epsscale{.80}
\plotone{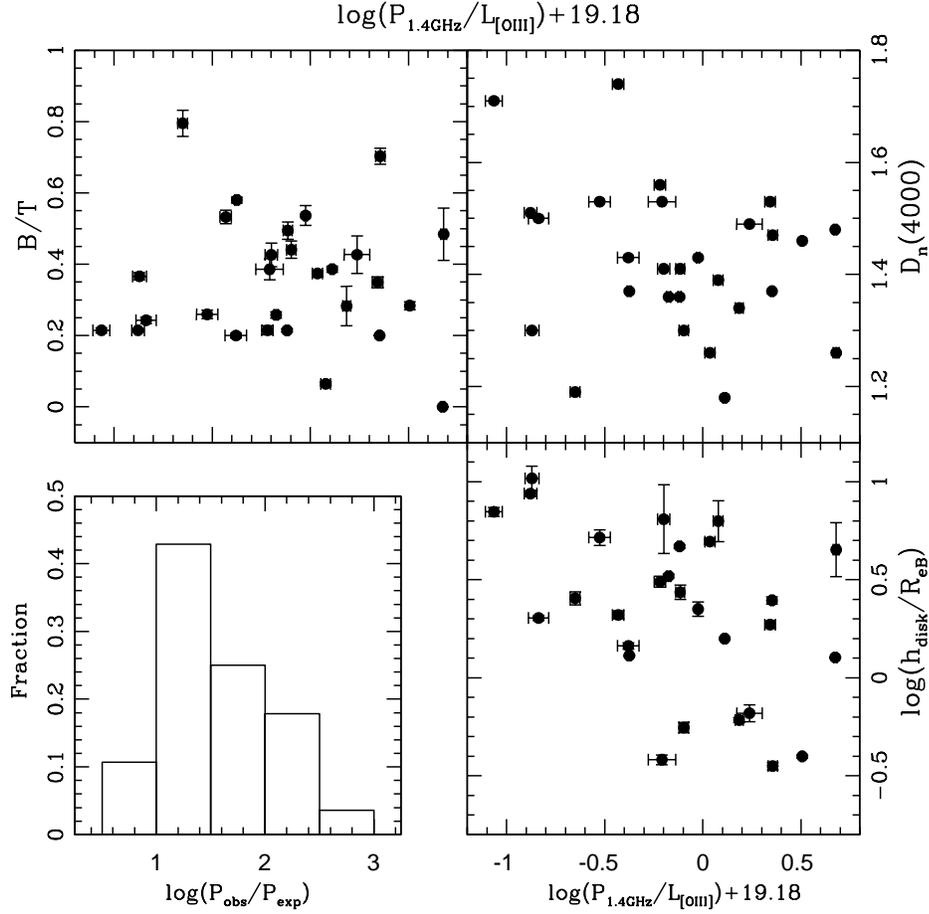}
\caption{The estimated radio loudness is plotted as a function of bulge ratio B/T (top-left panel), host stellar population
age $D_n(4000)$ (top-right panel) and ratio of the disk scale length over the bulge effective radius $h_d/r_e$ (bottom-right panel).
The isolated panel at the bottom-left corner shows the distribution of the ratio of the observed radio power over the 
expected maximum contributed by underlying star formations.
}
\end{figure}

\end{document}